\documentclass[prl,aps,twocolumn]{revtex4}

\pdfoutput=1
\usepackage{graphics}
\usepackage{graphicx}

\begin{document}

\title{Crystallization of the Lewis-Wahnstr\"{o}m ortho-terphenyl model}

\author{Ulf R. Pedersen}
\email{urp@berkeley.edu}
\affiliation{Department of Chemistry, University of California,
Berkeley, California 94720-1460, USA}
\author{Toby S. Hudson}
\email{hudson\_t@chem.usyd.edu.au}
\affiliation{School of Chemistry, University of Sydney, Sydney NSW 2006, Australia}
\author{Peter Harrowell}
\email{p.harrowell@chem.usyd.edu.au}
\affiliation{School of Chemistry, University of Sydney, Sydney NSW 2006, Australia}

\date{\today}

\begin{abstract}
Crystallization is observed during long molecular dynamics simulations of bent trimers, a molecular model proposed by Lewis and Wahnstr\"{o}m for ortho-terphenyl. In the crystal, the three spheres that make up the rigid molecule sit near sites of a body centered cubic lattice, the trimer bond angle being almost optimal for this structure. The crystal exhibits orientational disorder with the molecules aligned randomly along the three Cartesian axis (an example of cubatic orientational order). The rotational and translational mobilities exhibit only modest decreases on crystallization, by factors of 10 and 3 respectively. The rotational relaxation does change from Debye-like in the liquid to large angle jumps in the crystal. We consider the origin of the superior glass forming ability of the the trimer over the monatomic liquid.
\end{abstract}

\maketitle

\section{Introduction}
Molecular crystals, unlike their atomic counterparts, must accommodate the enormous variety of particle shape, symmetry and flexibility that molecules provide. This accommodation typically results in low symmetry crystals with a high propensity for polymorphism and twinning~\cite{silinish94,wright87}. There have been relatively few simulation studies of the crystallization of molecules from their melt and, of these, most have focussed on flexible polymers~\cite{polymer} or high symmetry molecules, either linear~\cite{linear} or octahedral~\cite{octah}. One notable exception is the freezing of water (a molecule with only $C_{2V}$ symmetry) which has, understandably, been the subject of a large number of studies~\cite{water}. With ordering dominated by directional hydrogen bonds, water is a special case and, therefore, offers little in the way of general insights into the nature of ordering of low symmetry organic molecules.

In this paper we report on molecular dynamics simulations of the crystallization of a liquid comprised of bent trimers (another $C_{2V}$ molecule), inspired by the organic glass-former ortho-terphenyl (OTP), in which the interactions are based on the Lennard-Jones potential. Since its introduction in 1955 by Andrews and Ubbelohde~\cite{andrews} as a useful liquid for the study of the glass transition, the properties of supercooled OTP have been the subject of extensive research~\cite{otpexpt}. In 1993, Lewis and Wahnstr\"{o}m (LW)~\cite{lewis93} introduced a model of OTP based on a rigid bent trimer of spherical particles which has subsequently been used for a number of simulation studies of glassy behaviour~\cite{lewis94,otpsim,otpsim_roskilde}.

Beyond the general similarity of the bent shape, the model bears little resemblance to OTP. The trimer angle, for example, was chosen to be $75^{\circ}$ rather than the $60^{\circ}$ value indicated by the structure of benzene because the authors regarded the former value, ``a value such that crystallization is prohibitively difficult''~\cite{lewis94}. The idea that it is possible to decide whether a molecule can crystallize easily or not (however `easily' might be defined here) simply by inspecting the shape of the molecule is, we suggest, both widespread and poorly justified. The crystallization of the trimer, as reported here, provides a salutary illustration of the stubborn ingenuity of periodic packing and the questionable value of our current intuition about how shape influences crystal stability and crystallization kinetics.

\section{Model and Algorithm}

The trimer is a rigid 3-particle complex, the site-site interactions all being a spherically symmetric Lennard-Jones potential $u(r)$ where $u(r)= 4\varepsilon\left(\left(\frac{\sigma}{r}\right)^{12}-\left(\frac{\sigma}{r}\right)^{6}\right)$ with $\varepsilon= k_B\times600\textrm{K } \simeq 4.988$ kJ/mol and $\sigma=0.483$ nm. The two trimer bonds are of length $\sigma$ and the fixed bond angle was chosen~\cite{lewis94} to be 75$^{\circ}$, presumably to avoid the choices  60$^{\circ}$ or 90$^{\circ}$, either of which might have been regarded as too easily accommodated within a face centered cubic lattice. (Although not really relevant to the subject of this paper, we note that \emph{real} bent trimer molecules must have bond angles between $\sim100^{\circ}$ and $120^{\circ}$. Ozone, with a bond angle of $116.78$~\cite{ozone-crystal}, has been proposed by Angell~\cite{ozone-glass} as a potentially good glass former based on the low value of its freezing point relative to its boiling point.) OTP, in contrast, has a bond angle 60$^{\circ}$ and, rather than spherical particles, it is composed of benzene rings that resemble oblate ellipsoids. Due to steric interactions (`overcrowding') the three benzene rings cannot be coplanar and, in the lowest energy configurations, the three rings lie on different planes. The crystal structure of OTP is orthorhombic (i.e different unit cell lengths along the three Cartesian axis) with 4 molecules per unit cell~\cite{otpcryst}, quite different, as we shall see, from the crystal structure of the LW trimer. More accurate models of OTP have been studied~\cite{accurate} in which each of the 18 C-H groups is represented as a spherical particle. These simulations are more time consuming than the LW model to run and, to date, no crystallization has been reported.

While the LW trimer may not provide a particularly realistic representation of OTP, it is of considerable interest as a computationally efficient and well studied model of a supercooled liquid comprised of a low symmetry molecule. We simulated a liquid of LW trimers using the \textsc{gromacs} software package \cite{gromacs}.  The tail of the Lennard-Jones interactions were smoothly brought to zero at $2.5\sigma$ using a shift function $S$: $U=\sum_{i>j}4\varepsilon\left(\left(\frac{\sigma}{r_{ij}}\right)^{12}-\left(\frac{\sigma}{r_{ij}}\right)^{6}-S\left(\frac{r_{ij}}{\sigma}\right)\right)$ where $S(r)=\gamma$ for $r<2.3$, $S(r)=\alpha(r-2.3)^3+\beta(r-2.3)^4+\gamma$ for $2.3>r>2.5$ and $S(r)=r^{-12}-r^{-6}$ for $r>2.5$ with $\alpha=0.2889\ldots$, $\beta=-0.7789\ldots$ and $\gamma=-0.005145\ldots$. Bonds were held rigid using the \textsc{lincs} algorithm \cite{hess1997}. Each Lennard-Jones force center were given a mass of 76.768 au giving a total molecular weight of 230.30 g/mol. Trajectories were evaluated for $N=324$ molecules in a periodic cubic box using a 8 fs time step for the integrator. Density and temperature were held fixed using the Nos{\'e}-Hoover thermostat with a time constant of 1 ps \cite{NoseHooverReferences}.

\section{Results}

\subsection{Spontaneous Crystallization of the Supercooled Liquid}

\begin{figure}
\begin{center}
\includegraphics[width=0.8\columnwidth]{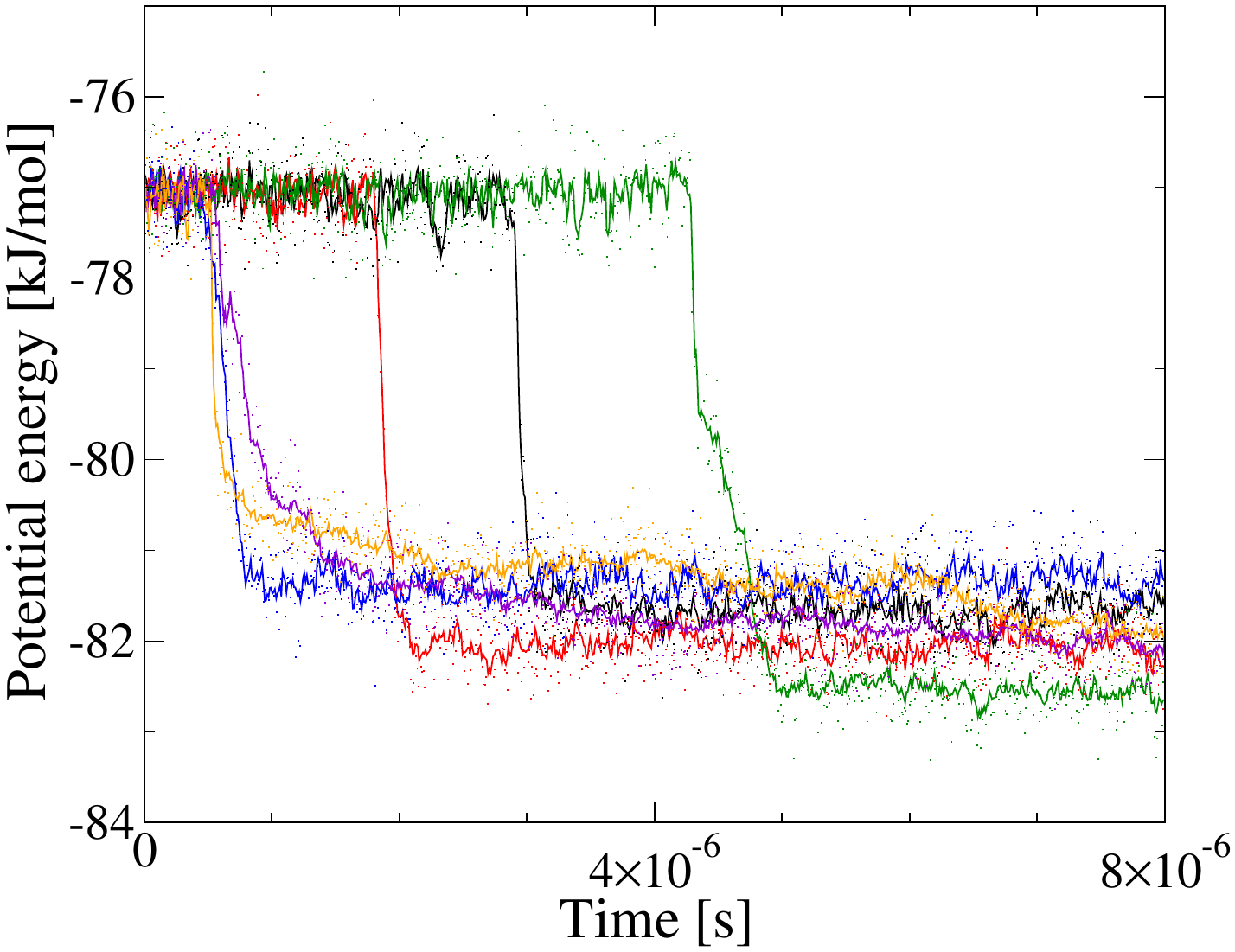}
\end{center}
\caption{\label{EpotOfRuns} Drop in potential energy due to crystallization of six independent runs at $\rho=1.135$ g/mL and $T=375$ K. The average crystallization time at this density, temperature and system size is estimated to $t_\textrm{cryst}=2\times10^{-6}$ s.}
\end{figure}

Simulations were performed at several temperatures along the zero pressure path, and along four density paths ($\rho=\{1.066,1.093,1.135,1.149\}$ g/ml) previously investigated \cite{otpsim_roskilde}. At sufficient supercooling, we observed an abrupt decrease in the potential energy from a stationary state (see Fig.~\ref{EpotOfRuns}) due to crystallization. The average lifetime $t_\textrm{cryst}$ of the supercooled liquid prior to crystallization at 1.135 g/mL and 375 K was $t_\textrm{cryst}=2\times10^{-6}$ s, a value roughly $10^2$ times that of $\tau_{\alpha}$, the structural relaxation time of the supercooled liquid.

\begin{figure}
\begin{center}
\includegraphics[width=0.48\columnwidth]{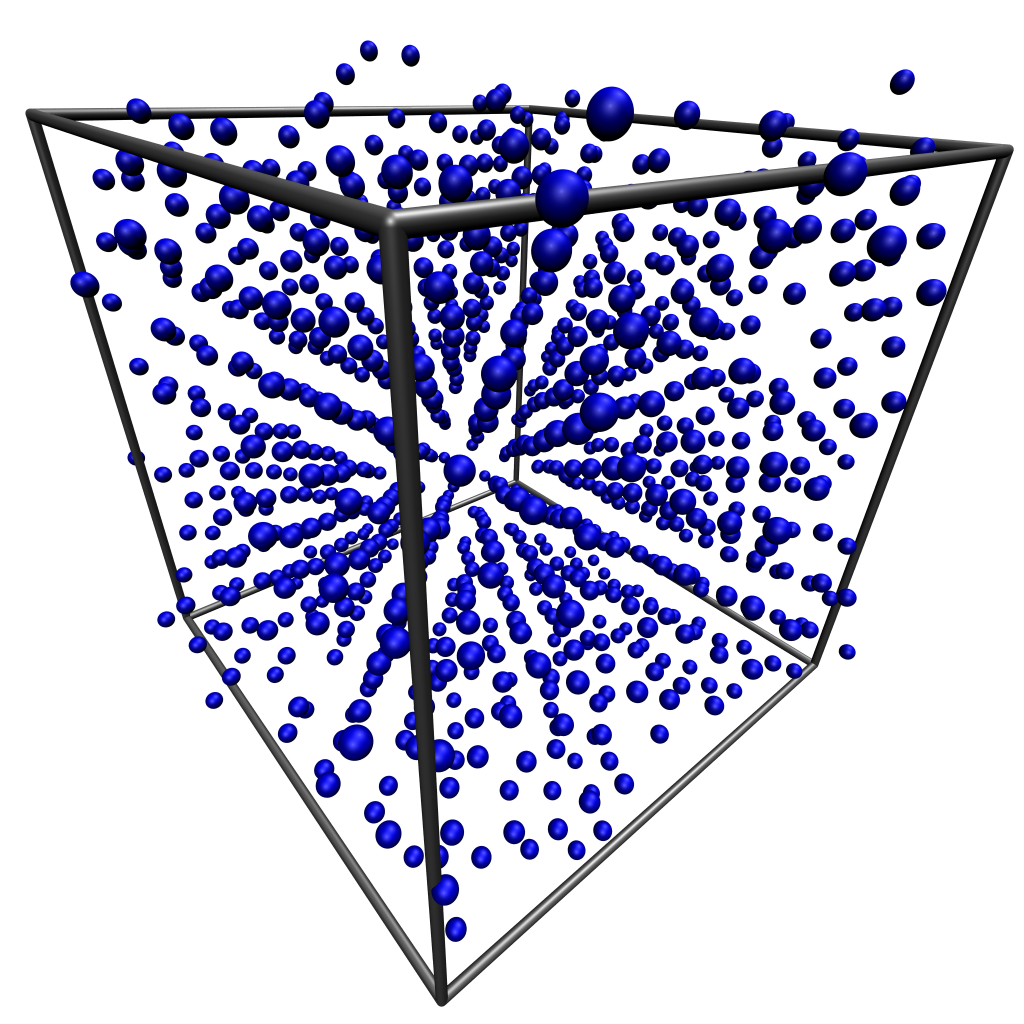}
\includegraphics[width=0.48\columnwidth]{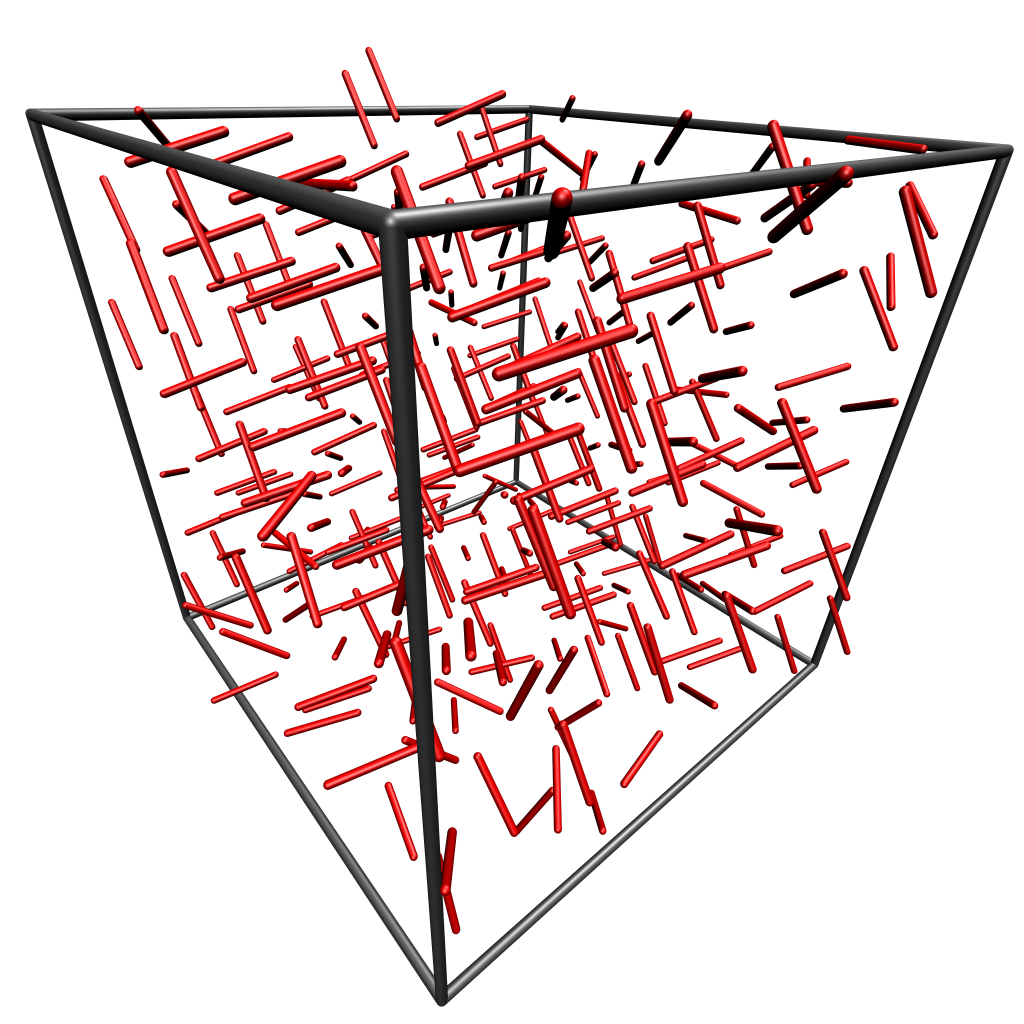}
\end{center}
\caption{\label{cryPic} Crystal configuration formed spontaneously from the supercooled liquid at $\rho=1.135$ g/mL and $T=375$ K. The left box display positions of the Lennard-Jones centers, that are in a near-BCC lattice. In the right box, only the line connecting the two outer atoms of the trimer are shown. We note the presence of considerable orientational disorder, constrained, however, to orientations that are either parallel or perpendicular to each other (i.e. cubatic order).}
\end{figure}

The crystalline order is clearly evident after the drop in potential energy as can be seen in the configuration, generated at the same conditions as those used in Fig.~\ref{EpotOfRuns}, depicted in Fig.~\ref{cryPic}. We also observed crystallization at \{250 K and 1.0937 g/mL; 260 K and 1.09272 g/mL; T=275 K and 1.09285 g/mL; 375 K and 1.14888 g/mL\}  (data not shown) in simulation runs of $\sim 10^{-6}$s duration.

\subsection{Crystal Structure}

\begin{figure}
\begin{center}
\includegraphics[width=0.4\columnwidth]{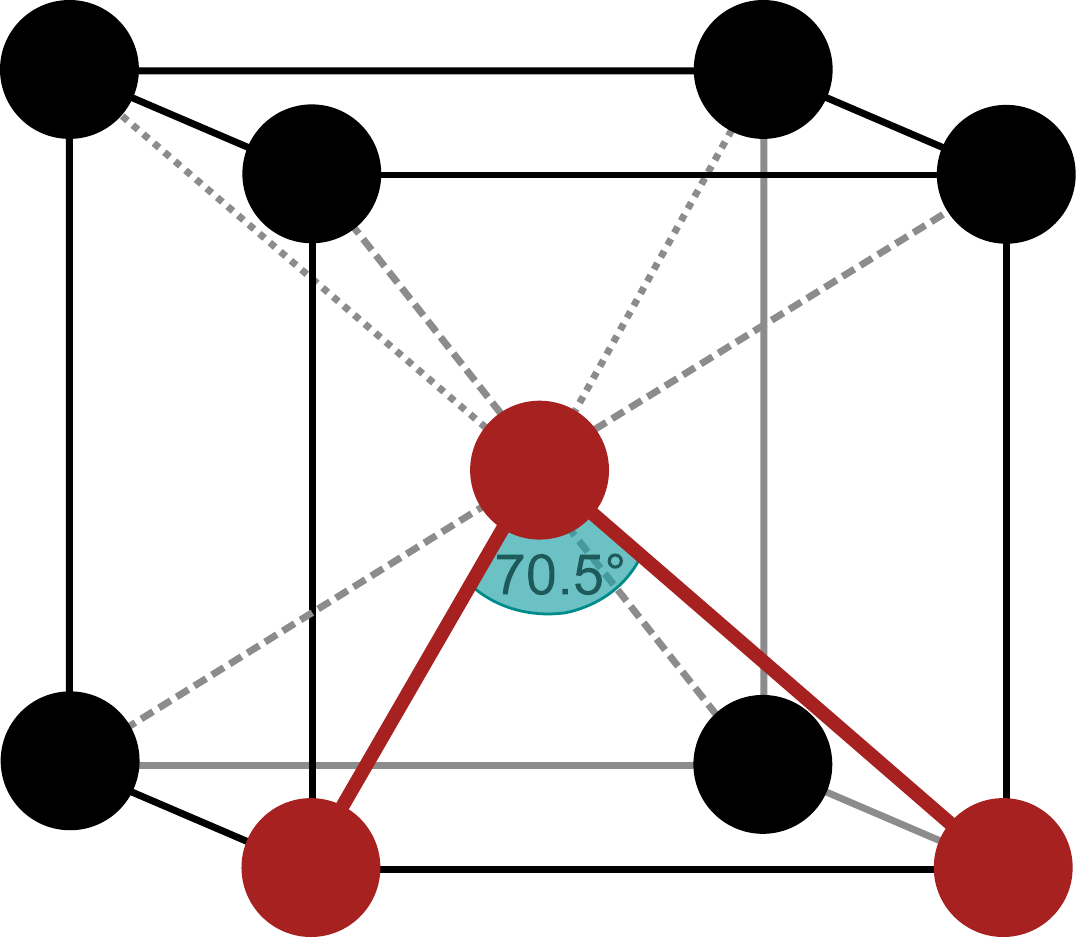}
\includegraphics[width=0.4\columnwidth]{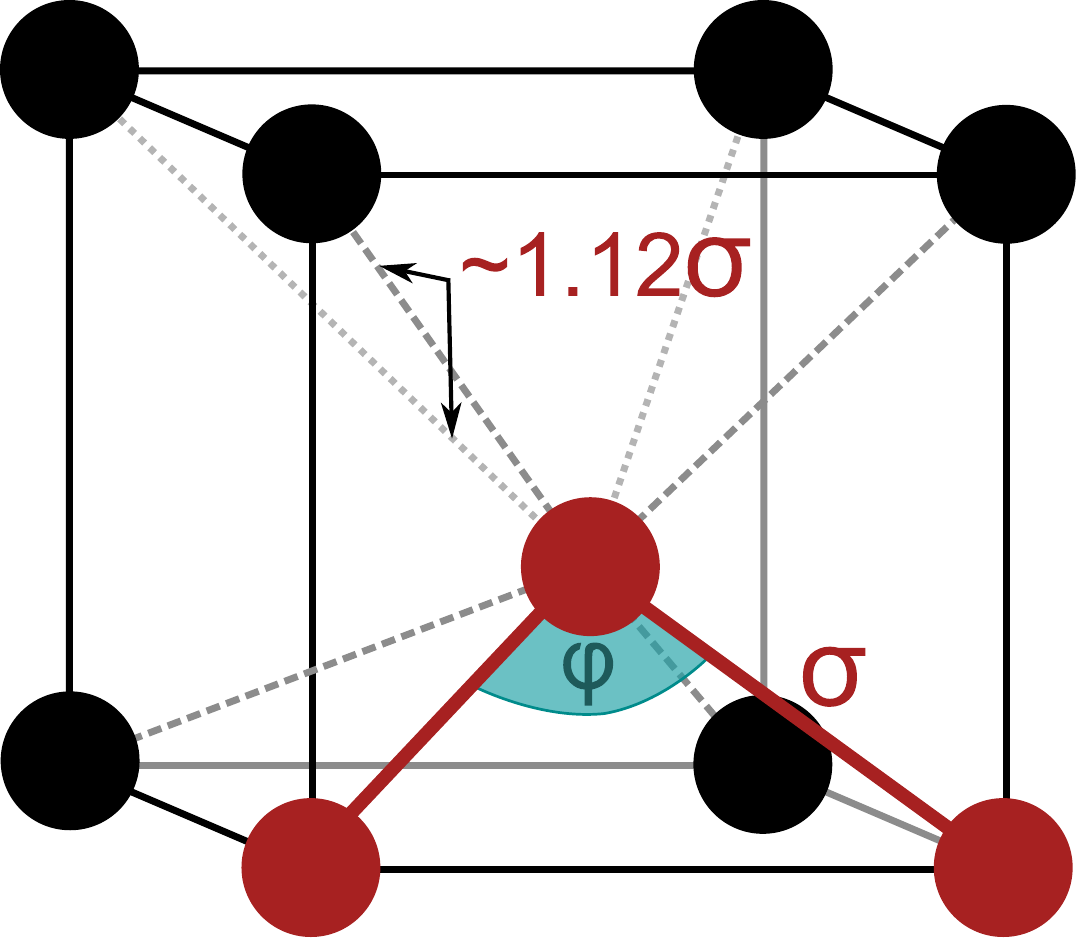}
\end{center}
\caption{\label{bcc} On the left, a trimer (indicated in red) with an optimal bond angle of $70.5^\circ$ occupies the sites of a BCC lattice. On the right, the trimer with a bond angle and bond length similar to that of the LW trimer is arranged on a near-BCC lattice.}
\end{figure}

\begin{figure}
\begin{center}
\includegraphics[width=.98\columnwidth]{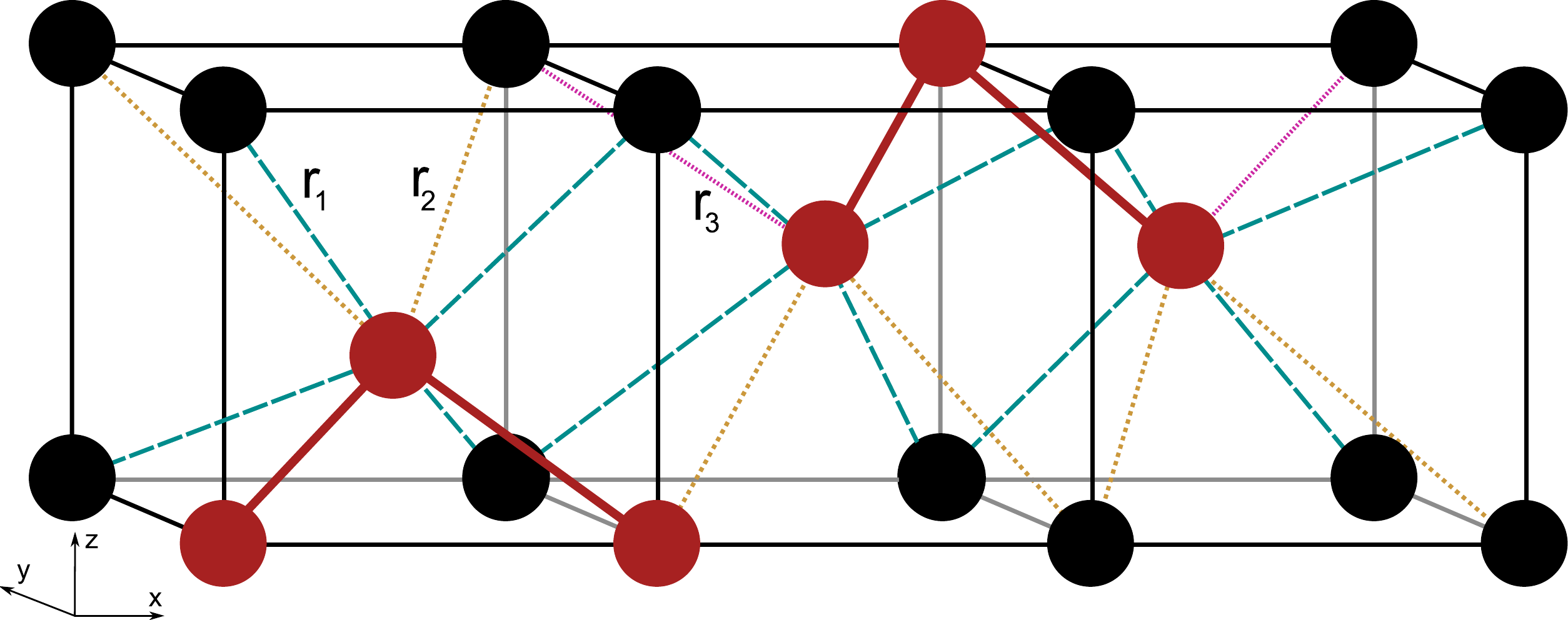}
\end{center}
\caption{\label{cell}The unit cell of an orientationally ordered arrangement of red trimers. Black sites represent the rest of the lattice sites, which are occupied by periodic images of the trimer sites. This forms a herringbone structure on the $(0\bar{1}1)$ plane. Nearest-neighbour interactions are marked. Note that this is an idealized structure missing some orientational freedom as discussed in the text.}
\end{figure}

Given the low symmetry of the LW trimer, the resulting crystal structure is surprisingly simple. The spherical particles that make up the trimer sit near the sites of a body centered cubic (BCC) lattice. To understand how this works, consider the idealized arrangements sketched in Fig.\ \ref{bcc}. Note that the smallest possible unit cell of the completely filled lattice contains two molecules, an example of which is shown in Fig. \ref{cell}.

\begin{figure}
\begin{center}
\includegraphics[width=0.8\columnwidth]{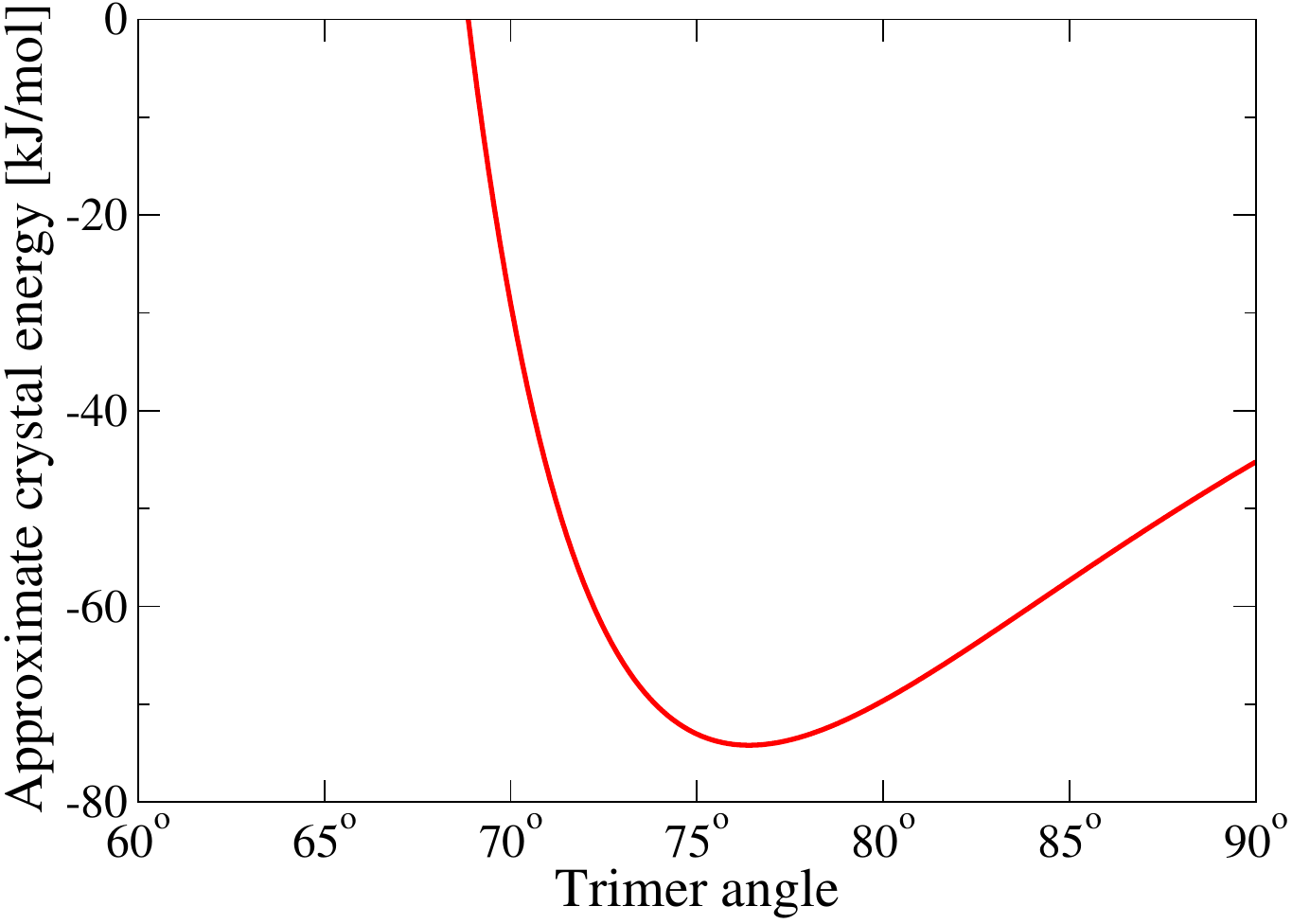}
\end{center}
\caption{\label{potential}. Potential energy of the crystal as estimated by Eq.~\ref{eq-pot} as a function of the trimer angle $\varphi$. The shift function $S$ use to truncate interactions smoothly in simulations introduce a $-72\varepsilon\gamma\simeq1.8$ kJ/mol shift of the shown energy.}
\end{figure}

The angle between bonds drawn from two adjacent corner sites to the centre is $2\sin^{-1}(1/\sqrt{3})\simeq70.5^\circ$. This would be the lowest energy choice for the bond angle {\it if} the bond lengths were increased from ${\sigma}$ to $1.12{\sigma}$ to reflect the fact that, in the BCC lattice, they must lie along the diagonal of the cubic cell. If we retain the bond length of $\sigma$, then the bond angle that minimizes the lattice energy will need to be larger than 70.5$^{\circ}$.

How far off is the LW trimer from the ideal bond angle for incorporation into the BCC lattice? To address this question we have calculated the approximate potential energy per molecule as a function of the trimer angle $\varphi$: Consider the orientationally ordered structure shown in Fig.\ \ref{cell} which, since it also fully decorates a near-BCC lattice, gives a good approximation to the interaction lengths in the actual crystal as formed in simulations.  The cell is constrained to retain the shape shown with three cubic subcells, the cubic edge length varies as $\varphi$ is varied, to ensure it corresponds to the distance between the end sites of the trimer.

The interactions of each Lennard-Jones site on the trimers with their 8 nearest-neighbours and 6 second-nearest neighbours were considered, apart from those defined by rigid bonds.  The 12 distances, $r_1$, represented in Fig.~\ref{cell} by long-dashed blue lines are slightly shorter than those represented by the 6 orange dotted lines, $r_2$.  The 2 magenta hashed lines have a fixed length of $r_3=\sigma$ due to the imposed requirement to remain cubic. There are 16 second-nearest interaction distances per two molecules (after removing rigid bonds and double counting), represented by solid lines which are the length of the cube edge, $a$, or a little longer for displaced sites. Thus the interaction energy per molecule is approximately
\begin{equation}
U_\textrm{cry}/N\simeq6u(r_1)+3u(r_2)+u(r_3)+8u(a)
\label{eq-pot}
\end{equation}
where $a/\sigma=\sqrt{2-2\cos(\varphi)}$, $d/\sigma=\frac{\cos(\frac{\varphi}{2})}{\sqrt{2}}$, $r_1=\sqrt{(\frac{a}{2})^2+d^2+(a-d)^2}$, and $r_2=\sqrt{(\frac{a}{2})^2+2(a-d)^2}$.
The estimated crystal energy, shown in Fig.\ \ref{potential}, is minimized at $\varphi=76.4^\circ$ corresponding to a density of 0.352 molecules per $\sigma^3$ or 1.19 g/mL. This suggests that the LW trimer angle of 75$^\circ$ is, in fact, a near-ideal geometry to crystallize into this dense, low-energy crystal structure.

While the constitutive atoms are quite comfortably accommodated on the BCC lattice, the trimer orientation exhibits considerable disorder. The orientation of a trimer can be established by the vector parallel to the line joining the two outer atoms. In the right panel on Fig.\ \ref{cryPic} we show one configuration of these orientational vectors in the crystal (where we have omitted the sense of the vector). In contrast to the regular arrangement of constituent spheres, the orientational vectors exhibit considerable disorder. The disorder is not completely random since restrictions are imposed by the BCC ordering of the spheres.

\begin{figure}
\begin{center}
\includegraphics[width=0.8\columnwidth]{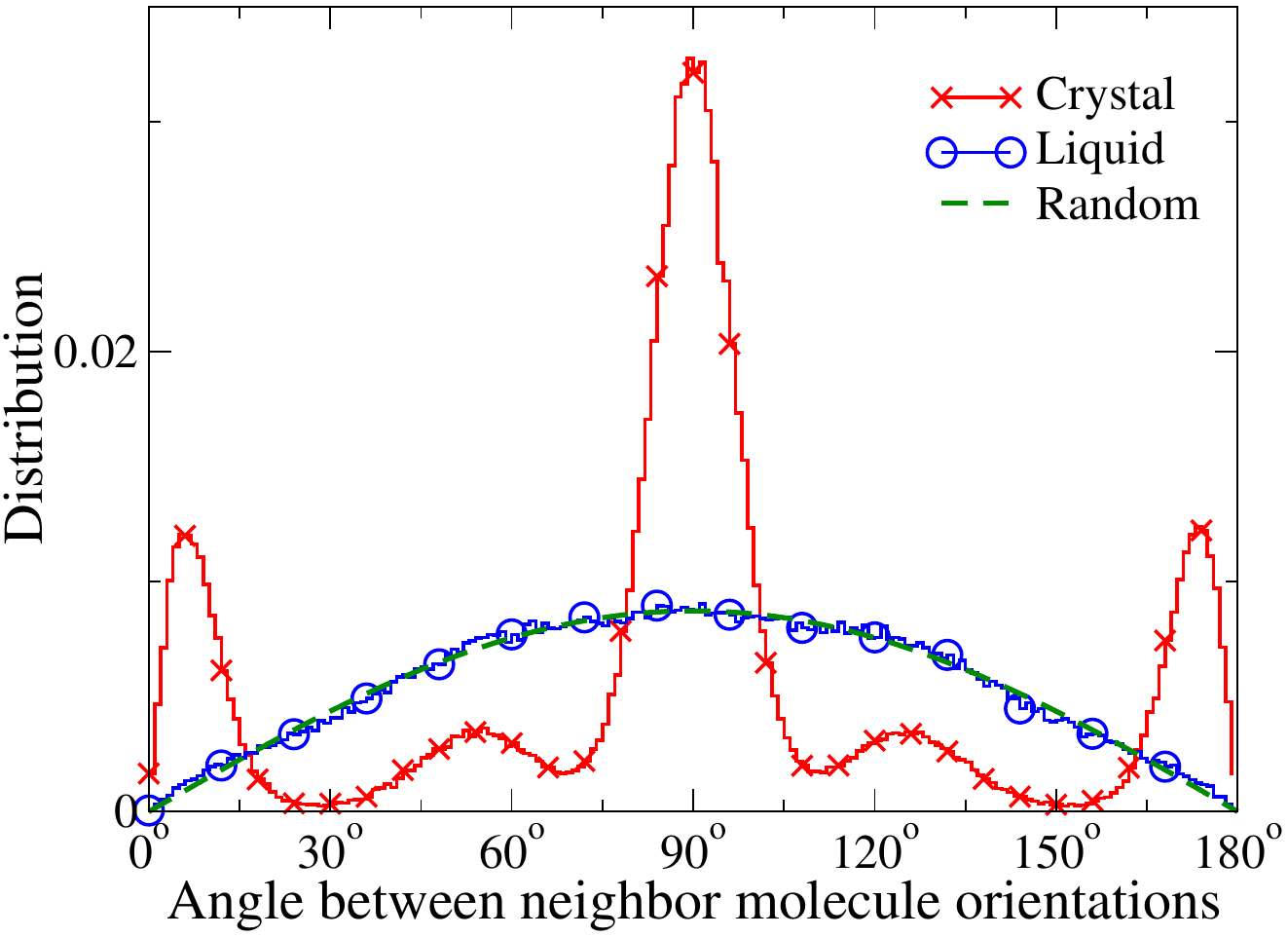}
\end{center}
\caption{\label{angleHist} Distribution of the angles between the orientation vectors of neighbors in the crystal (as generated by simulation) and in the liquid. The green dashed line indicates a random distribution.}
\end{figure}

The result is a form of orientational order known as {\it cubatic} \cite{cubatic} in which the orientation vectors (as we have defined it) are either parallel or perpendicular. This restriction is evident in the peaks in the distribution of angles between pairs of orientation vectors at $0^\circ$, $90^\circ$ and $180^\circ$ shown in Fig.~\ref{angleHist}. The smaller peaks at $\sim 55^\circ$ and $125^\circ$ correspond to orientational defects. The appearance of this cubatic order during crystallization can be monitored through an order parameter defined as
\begin{equation}
Q_i=\frac{1}{N-1}\sum_{j\neq i}^N \frac{\cos^2(2\theta_{ij})-\frac{7}{15}}{1-\frac{7}{15}}
\label{Qi}
\end{equation}
where $\theta_{ij}$ is the angle between the orientation vectors of molecule $i$ and $j$ and $\frac{7}{15}$ is the integral of $\cos^2(2\theta_{ij})$ over a random distribution of orientations. A global cubatic order parameter $Q$ is defined as the average $Q_i$. Note that $Q$ is normalized so that zero correspond to random orientations, and unity to perfect cubatic ordering.

\begin{figure}
\begin{center}
\includegraphics[width=0.8\columnwidth]{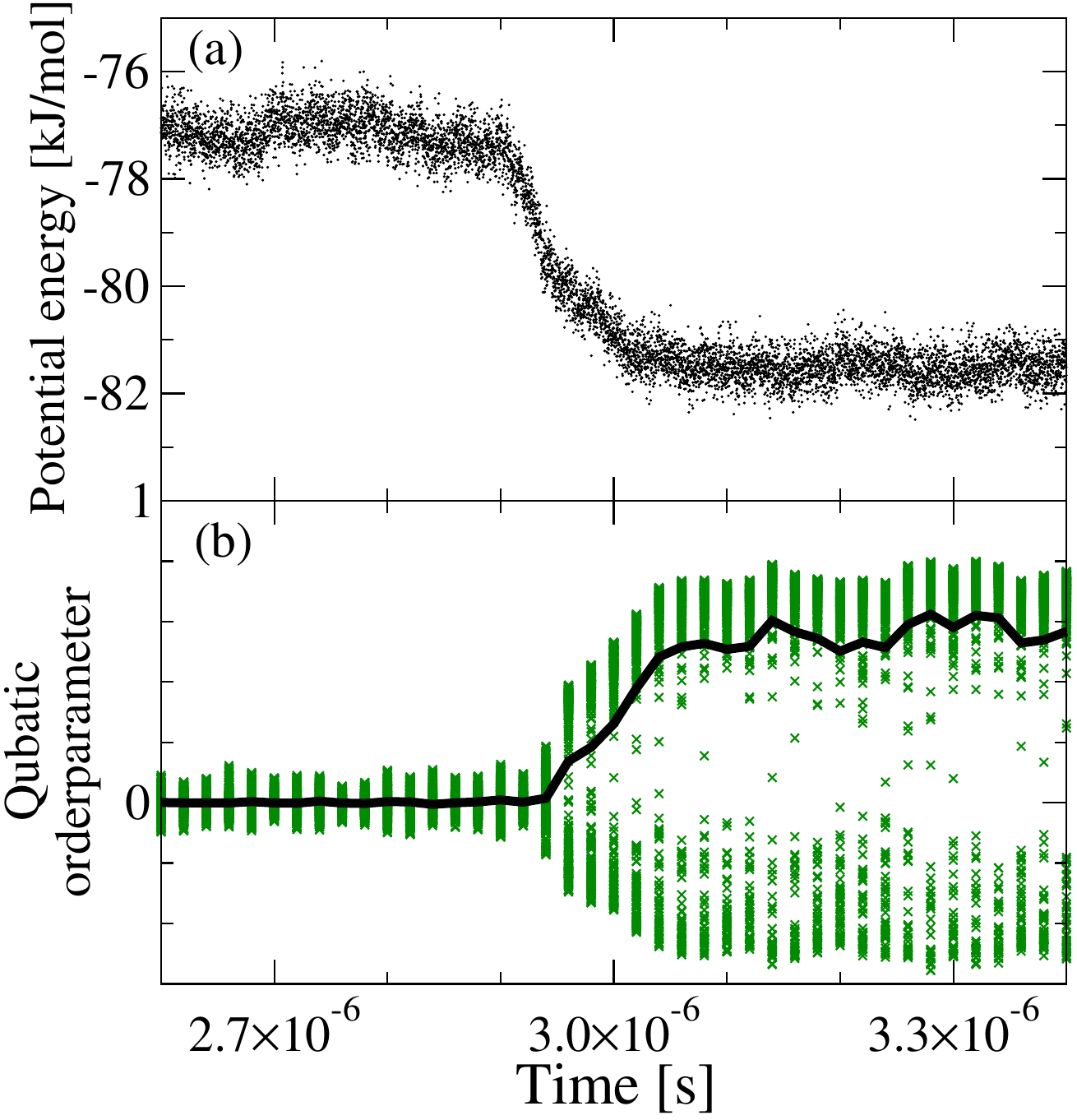}
\end{center}
\caption{\label{EpotQ} The drop in potential energy (a) is accompanied by cubatic ordering (b). Green crosses indicates cubatic order parameter $Q_i$'s (Eq. \ref{Qi}) and the black line the global cubatic order parameter $Q$ ($\rho=1.135$ g/mL, $T=375$ K). Negative $Q_i$'s in the crystalline phase are associated with orientational defects .}
\end{figure}

The cubatic order parameters along a crystallizing trajectory are shown on Fig.\ \ref{EpotQ}. We find that the increase in the cubatic order follows closely the decrease in the potential energy of the inherent structures during the course of crystallization. In particular, we can see no sign of a development of the orientational order preceding the onset of crystallization.

\subsection{Relaxation Kinetics in the Liquid and Solid Phases}

\begin{figure}
\begin{center}
\includegraphics[width=0.75\columnwidth]{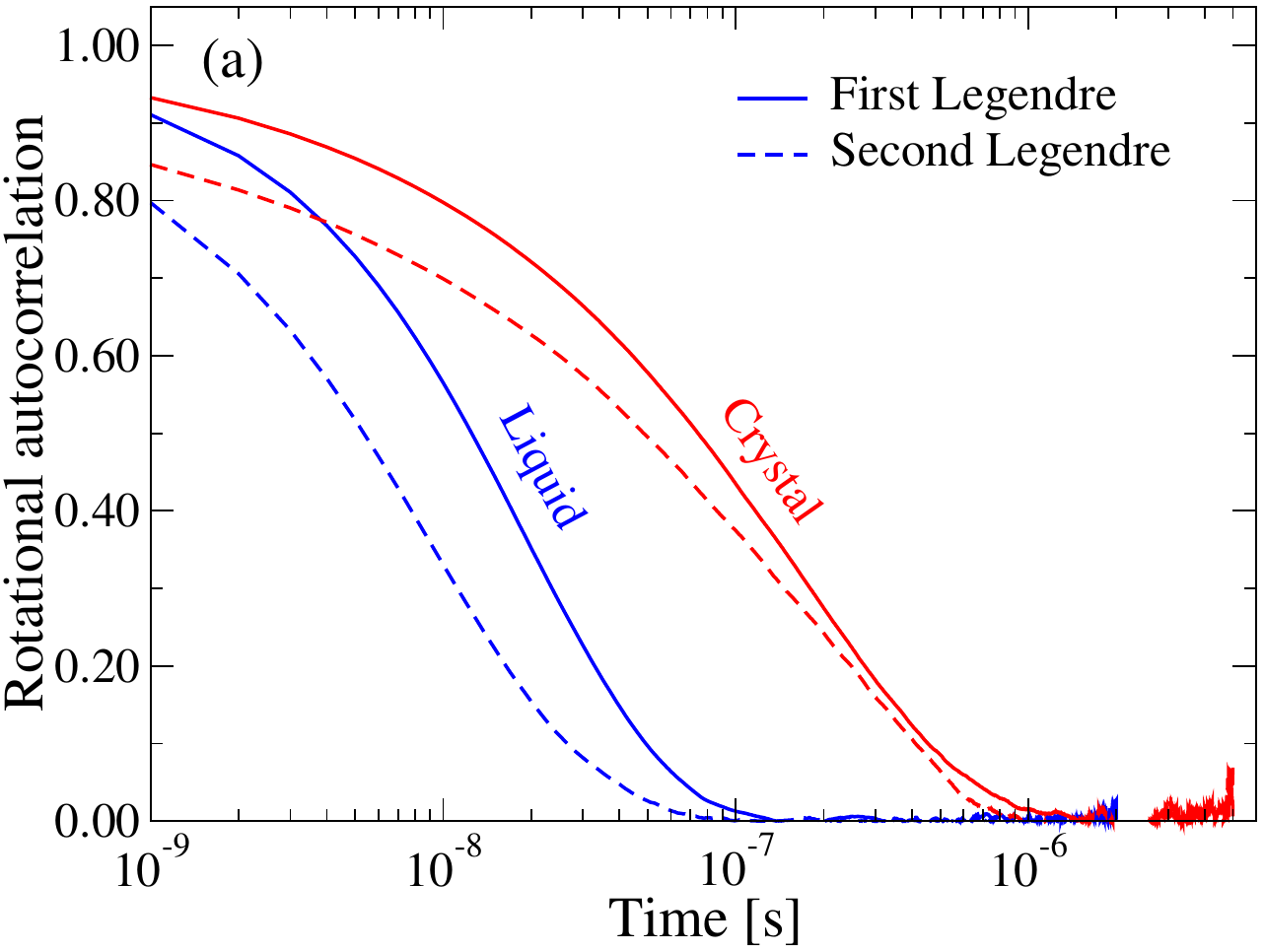}
\includegraphics[width=0.75\columnwidth]{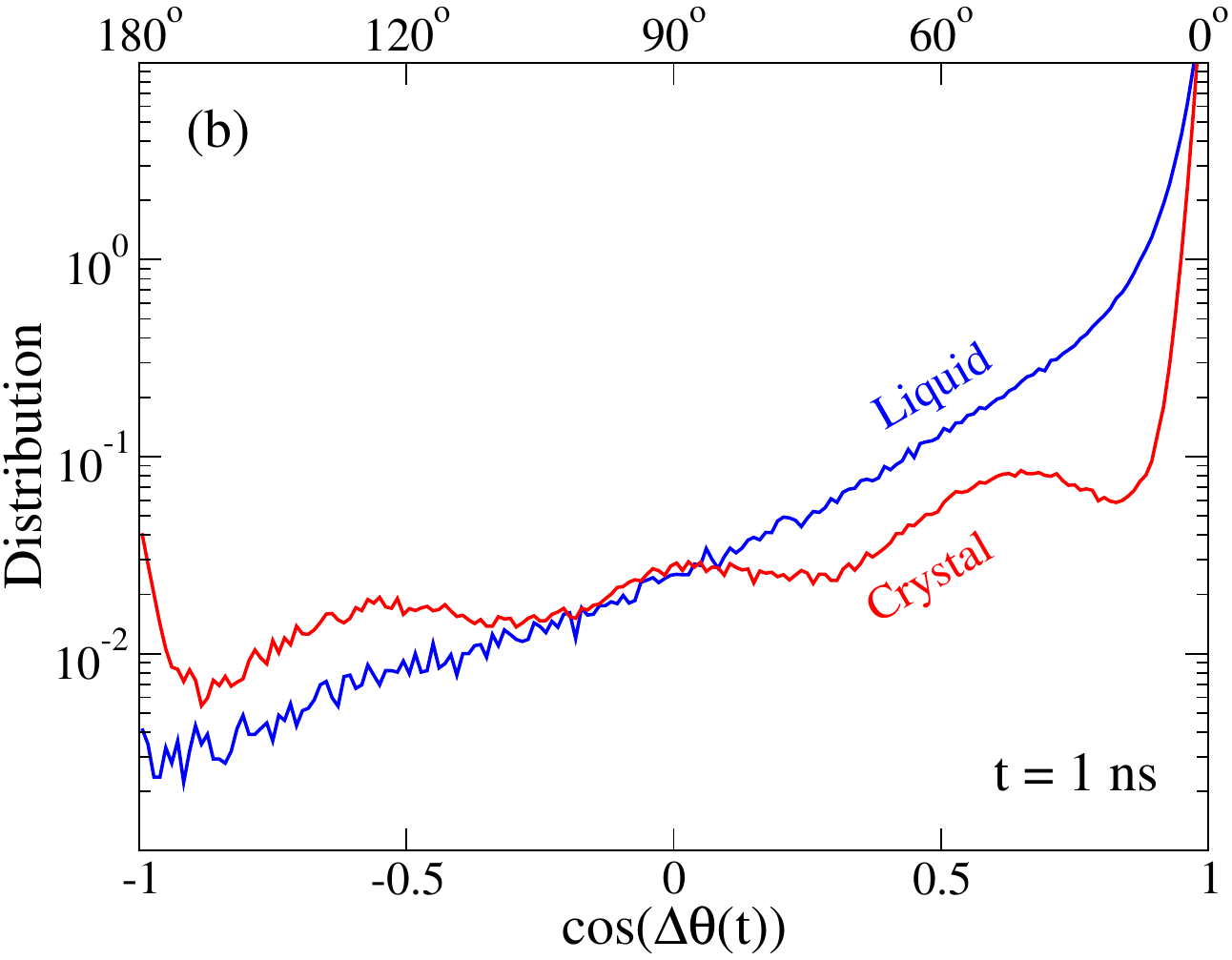}
\end{center}
\caption{\label{RotRelax} (a) Rotational autocorrelation functions defined as $C_l(t)=\langle P_l\{\cos[\theta(0)]\} P_l\{\cos[\theta(t)]\} \rangle$ where $\theta$ is the angle between the trimer base and either $\vec{x}$, $\vec{y}$ or $\vec{z}$ and $P_l$ is the Legendre polynomial of either rank $l=1$ (full lines) or $l=2$ (dashed lines). $C_l(t)$'s are evaluated for both the liquid (blue) and crystal  (red) at $\rho=1.135$~g/mL and $T=375$ K. Rotational relaxation time is only affected by a factor of $\sim 10$ on crystallizing. (b) Distribution of angular displacements of molecular orientations after 1 ns.}
\end{figure}

In light of the disorder retained by the trimer crystal, how mobile are the molecules in the crystal? In Fig.\ \ref{RotRelax}(a) we plot the autocorrelation function for the angle between a molecule's orientation vector at time $t = 0$ and the vector at some later time $t$ for the liquid and the crystal phases. We find that rotational relaxation persists into the crystal phase and the rotational relaxation time increases, on freezing, by only a factor of 10.

The ratio of the relaxation times of the Legendre polynomials of rank 1 and 2 is $\sim 2$ in the liquid and $\sim 1$ in the crystal, indicating that the rotational relaxation changes on freezing from Debye-like diffusion in the liquid to relaxation involving large angular jumps in the crystal~\cite{kivelson}. In Fig.~\ref{RotRelax}(b) we have plotted the distribution of angular displacements after a short time ($t = 1$ ns, roughly $5 \%$ of the liquid rotational relaxation time) in the liquid and crystal phases at density $1.135$ g/mL and temperature $375$ K. While the liquid shows the smooth distribution expected for incremental displacements, the distribution of angular jumps in the crystal exhibits a number of gentle peaks. The predominant peak does not occur at an angle of $90^\circ$ or $180^\circ$, as one expect if the jumps where from on crystal-aligned position to another, but, rather, at an angle of $\sim 55^\circ$. We conclude that rotational relaxation in the crystal involves orientational defect configurations as transient intermediates.

\begin{figure}
\begin{center}
\includegraphics[width=0.8\columnwidth]{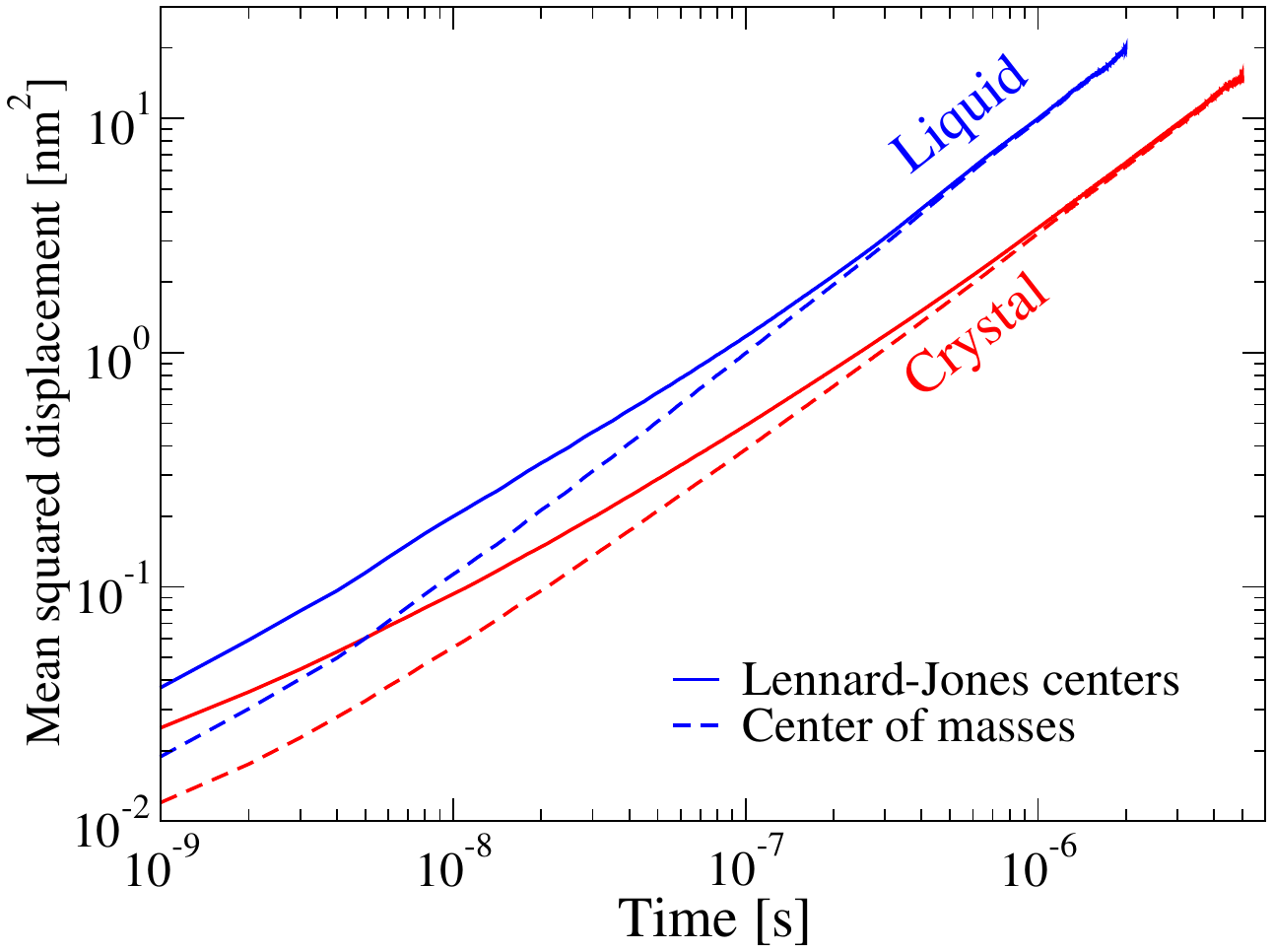}
\end{center}
\caption{\label{MSD} Mean squared displacement of LJ centers (full lines) and center of masses (dashed lines) in the crystal  (red) and liquid (blue) phases at $\rho=1.135$ g/mL and $T=375$ K. Center of mass diffusion constants
in liquid and crystal, estimated from long time displacements, are $1.6 \times 10^{-12}$~m$^{2}$/s and $5.3 \times 10^{-13}$~m$^{2}$/s, respectively.}
\end{figure}

The mean square displacement, plotted in Fig.\ \ref{MSD}, shows a similar picture of only a modest slowing down on crystallization. The centre of mass motion of the trimer in the crystal continues to exhibit diffusion-like behaviour at long times. At $T=375$K the center of mass diffusion constants for the liquid and crystal are $1.6 \times 10^{-12}$~m$^{2}$/s and $5.3 \times 10^{-13}$~m$^{2}$/s, respectively, corresponding to a slow down of only a factor of three.  These calculations have been performed at a finite fixed volume and there will be a pressure drop occurring at crystallization. Even taking this into account, the persistent mobility of the trimers in the crystal is striking.

We find that the mobility in the crystal is sensitive to the amount of orientational defects present and that, in some runs, this defect density decreased through an aging process after crystallization. Specifically, in four of the crystals formed at $\rho=1.135$ g/mL and $T=375$ K (see Fig.~\ref{EpotOfRuns}) the amount of orientational defects fluctuates around 10\%, while in the other two runs, we observe aging of the crystal, characterized by an increase in the cubatic order parameter. The relaxation times for translational and rotational diffusion are observed to increase with the increase in orientational order. We emphasize, however, that the observed crystallization in all runs involves a transition to the crystals with $\sim 10\%$ orientational defects.

\section{Discussion}
In this paper we have shown that the Lewis and Wahnstr\"{o}m bent trimer will crystallize. The resulting structure is a plastic crystal in which rotational motion and translational diffusion persist at rates only modestly reduced from those of the liquid. While the observation of crystallization contradicts the model designers' expectation that crystallization of the trimer would prove ``prohibitively'' difficult, we find that their intuition that the trimer is, in some sense, difficult to freeze is still correct. To see this, consider the \emph{minimum} crystallization time $t_{cryst}^ {min}$ for the LW trimer and the atomic Lennard-Jones liquid. ($t_{cryst}^{min}$ corresponds to the `nose' in the standard time-temperature transformation plot.) For the atomic liquid (at a density of  $\rho=1.135$ g/mL) we find $t_ {cryst}^ {min} = 3 \times 10^{-10}$s while the smallest values of $t_ {cryst}$ obtained so far for the LW trimer is $\sim 2 \times 10^{-6}$s. The factor of $10^{4}$ difference in these two values of $t_ {cryst}^ {min}$ corresponds to a substantial increase in the glass forming ability of the trimer over that of the atomic liquid.

How we account for this increase represents an important puzzle. It has been suggested that slow crystallization is a consequence of `bad' (read `high enthalpy') crystals~\cite{angell}. The idea is attractive as it reduces the problem of adjusting glass forming ability to one of adjusting the lattice energy of the stable crystal.  A direct measure of crystal stability is the change in energy on freezing relative to the energy of the metastable liquid, i.e. $\Delta U_{cryst}/U_ \textrm{liq}$. At the temperatures associated with the respective peak crystallization rates, we find $\Delta U_ {cryst}/U_ {liq}= 0.058$ for the trimer and $0.089$ for the liquid of LJ atoms.

This preliminary comparison supports the proposition that the improved glass forming ability of the trimer \emph{might} be a result of the decreased stability its crystal, relative to that of the atomic crystal for which shape is not an issue. To establish whether this stability difference is sufficient to account for the trimer's glass forming ability requires additional analysis that we leave for a future paper.

\section{Acknowledgments}
URP is supported by The Danish Council for Independent Research in Natural Sciences. TSH and PH acknowledge support from the Australian Research Council. We gratefully acknowledge computation time provided by the Danish National Research Foundation Center ``Glass and Time''.

\begin{center}
\includegraphics[width=.8\columnwidth]{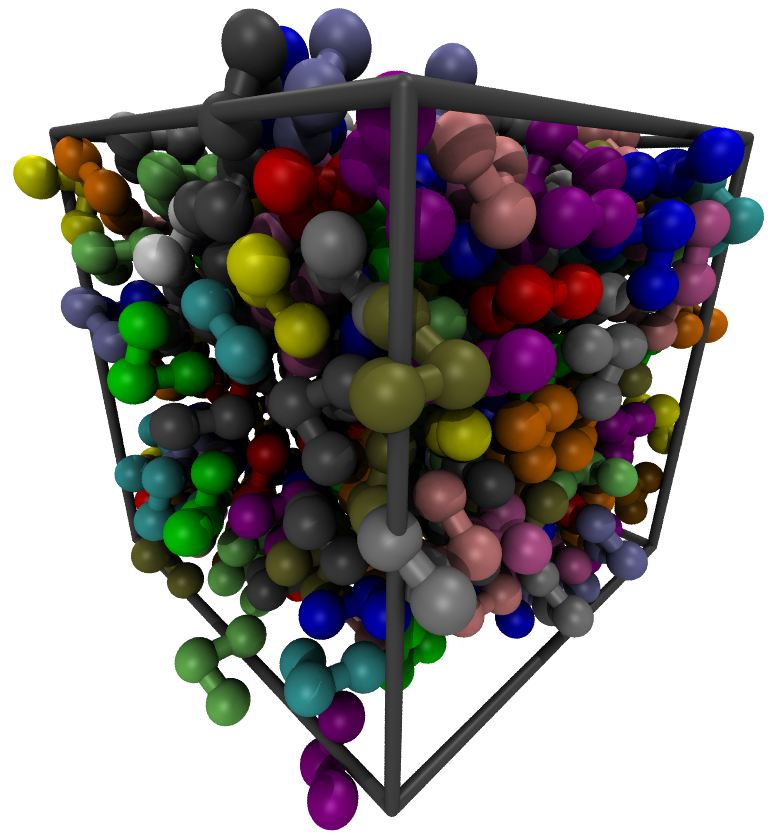}\\
\end{center}
Supplementary figure: The Lewis-Wahnstr{\"o}m ortho-terphenyl crystal.

\end{document}